\documentclass[fleqn,usenatbib]{mnras}

\usepackage[utf8]{inputenc} 
\usepackage[T1]{fontenc}    
\usepackage{hyperref}       
\usepackage{url}            
\usepackage{booktabs}       
\usepackage{amsfonts}       
\usepackage{nicefrac}       
\usepackage{microtype}      
\usepackage{lipsum}

\title{Effectively Calculating Gaseous Absorption in Radiative Transfer Models of Exoplanetary and Brown Dwarf Atmospheres}

\usepackage{newtxtext,newtxmath}

\usepackage[T1]{fontenc}
\usepackage{ae,aecompl}

\usepackage{graphicx,siunitx}	
\usepackage{amsmath,textcomp}	
\usepackage{amssymb}	
\usepackage[version=3]{mhchem}
\usepackage{multirow,rotating,adjustbox,pdflscape,hyperref}

\author[Garland \& Irwin]{
Ryan Garland,$^{1}$\thanks{E-mail: ryan.garland@physics.ox.ac.uk}
Patrick G. J. Irwin,$^{1}$
\\
$^{1}$Department of Physics, University of Oxford, Oxford, United Kingdom 
}

\date{Accepted XXX. Received YYY; in original form ZZZ}

\pubyear{2019}

\begin{document}
\label{firstpage}
\pagerange{\pageref{firstpage}--\pageref{lastpage}}
\maketitle

\begin{abstract}

Sophisticated atmospheric retrieval algorithms, such as Nested Sampling, explore large parameter spaces by iterating over millions of radiative transfer (RT) calculations. Probability distribution functions for retrieved parameters are highly sensitive to assumptions made within the RT forward model. 
One key difference between RT models is the computation of the gaseous absorption throughout the atmosphere. We compare two methods of calculating gaseous absorption, cross-sections and correlated-$k$, by examining their resulting spectra of a number of typical \ce{H2}-He dominated exoplanetary and brown dwarf atmospheres. We also consider the effects of including \ce{H2}-He pressure-broadening in some of these examples. 
We use NEMESIS to compute forward models. Our $k$-tables are verified by comparison to ExoMol cross-sections provided online and a line-by-line calculation. 
For test cases with typical resolutions ($\Delta \nu = 1$cm$^{-1}$), we show that the cross-section method overestimates the amount of absorption present in the atmosphere and should be used with caution. For mixed-gas atmospheres the morphology of the spectra changes, producing `ghost' features. The two methods produce differences in flux of up to a few orders of magnitude. The addition of pressure broadening of lines adds up to an additional order of magnitude change in flux. These effects are more pronounced for brown dwarfs and secondary eclipse geometries. We note that correlated-$k$ can produce similar results to very high-resolution cross-sections, but is much less computationally expensive.
We conclude that inaccurate use of cross-sections and omission of pressure broadening can be key sources of error in the modelling of brown dwarf and exoplanet atmospheres.

\end{abstract}




\section{Introduction}

The Solar System contains a wonderful variety of planetary atmospheres. Each planet has its own unique formation and evolutionary history resulting in brilliant rings, icy moons, and complex atmospheric compositions. In order to better understand the physical processes taking place within these celestial bodies, we extract data across almost the entire electromagnetic spectrum, primarily in the visible and infrared as these are the wavelengths where light is best reflected and thermally emitted. Radiative transfer models allow us to simulate the physical processes on these bodies, producing spectra that can be compared to our data sets. Usually these radiative transfer models are coupled to an inverse method, which iteratively explores parameter space to retrieve the most likely set of parameters describing the atmosphere of the (exo)planet or brown dwarf (e.g., \citet{nemesis08, waldmann15, line15}). 

This method of determining the atmospheric structure and composition is reliant on the accuracy of our radiative transfer models. For Bayesian retrieval algorithms such as Nested Sampling (e.g. MultiNest \citet{feroz2013}) and Markov Chain Monte Carlo (MCMC, e.g. EMCEE \citet{foreman-mackey13}), the radiative transfer forward model will be executed millions of times while it explores a large parameter space. Because of the nature of these highly dimensional retrieval calculations, small differences in forward models could potentially produce non-trivial changes in the retrieved probability distributions. For example, if one model produces an extra, erroneous absorption band for, say, \ce{NH3}, the retrieval would try to compensate for this error during the fitting procedure. While the `true' value of \ce{NH3} should fit the spectrum (assuming the rest of the model is perfect), the extra absorption band would mean that the total amount of retrieved \ce{NH3} will be smaller than the `true' value, as it finds a medium-ground answer which best fits all (true + erroneous) absorption bands. If we do not assume the rest of the model is perfect, then the problem is even graver - the retrieval will try to fit the spectrum by adjusting the model in a non-trivial way, such as adjusting the surface gravity or the temperature profile. In this way, one forward model error propagates through to all other retrieved posterior probability distribution functions (PDFs).

A key difference between a number of radiative transfer models is how the gaseous absorption is calculated - this is because it is a computationally expensive step. A line-by-line calculation, where the absorption coefficient is calculated for the exact temperature and pressure for each individual line, is the most accurate and correct as it resolves each individual line exactly. However, this method is far too slow to be used in sophisticated retrieval procedures because of the enormous number of lines involved, especially at high temperatures (>1000K). Instead, the absorption coefficient is usually pre-calculated at a range of temperatures and pressures for a variety of gases, and interpolated to the necessary values. There are multiple methods of calculating the gaseous absorption, but here we limit ourselves to discussion of just two methods: individual gas cross-sections and correlated-$k$ \citep[e.g.,][]{lacis1991}. We also briefly consider the opacity sampling method \citep[see e.g.][]{hubeny2014}, though not as extensively as the previous two as this would require a major overhaul of our code. We note that, as long as the pressure and temperature resolution in the cross-section look-up table is high enough, there is no distinction between a high-resolution ($\Delta\nu = 0.01$ cm$^{-1}$) cross-section and a line-by-line calculation.
Individual gas cross-sections have been used to solve the radiative transfer equation \citep{macdonald2017}. The cross-sections are absorption coefficients calculated from a line-by-line calculation on a grid of pressures and temperatures for a given gas, then integrated preserving area to a (typical) resolution of $\Delta \nu = 1$cm$^{-1}$. Other authors calculate a high-resolution cross-section and sample it at a resolution of $\Delta \nu = 1$cm$^{-1}$ \citep{line15, sharp07}. \citet{hedges2016} found that these individual gas cross-sections could have median differences of <1\% for low-resolution ($R\sim$ 100), up to 40\% for medium-resolution ($R\lesssim$ 5000), and over 100\% to 1000\% for high-resolution cross-sections ($R\sim$ \num{1e5}), introduced by various aspects of
pressure broadening. This is before the non-trivial task of quantifying these differences over a whole atmosphere.

Another approach to calculating the spectra of brown dwarfs and exoplanets is using premixed $k$-coefficients \citep{saumon08} or on-the-fly mixing for the correlated-$k$ method \citep{nemesis08, barstow14, jaemin14}. Note that for the rest of the paper, `cross-sections' refers to cross-sections which are not premixed, i.e. individual cross-sections for individual gases. $k$-tables are produced by performing line-by-line calculations of the absorption coefficient then rewriting the absorption coefficient strength distribution in terms of a cumulative frequency distribution over bins of specified wavenumber/wavelength width by ranking and sampling the distribution according to absorption coefficient strength. The inverse of this distribution is known as the $k$-distribution \citep{lacis1991}. The $k$-distribution is a smooth, monotonically increasing function and so can be sampled with only 10-20 points, compared with $\sim$ $10^3$ to $10^6$ for the cross-section or line-by-line methods. Within a single atmospheric layer, we may simply combine the $k$-distributions of different gases by assuming that the lines are randomly overlapping. Note that there are many ways of combining $k$-coefficients for different gases. Among brown dwarf and gas giant atmosphere models, ATMO \citep{tremblin2015,drummond2016} and PETIT \citep{molliere2015} also use random overlap, HELIOS \citep{malik2017} assumes perfect correlation, and \citet{amundsen2016} use equivalent extinction. For further information see \citet{amundsen2017}.  Then, by assuming that the wavenumbers at which the (total) cross section takes a certain value are vertically correlated these $k$-distributions may be used to calculate the transmission, thermal emission or scattering of an atmosphere using the correlated-$k$ method. From previous studies \citep{nemesis08}, we have found the correlated-$k$ approximation to be accurate to better than 5\%. This is why we primarily use this as our benchmark absorption method for the paper. For more information on our correlated-$k$ method, refer to \citep{nemesis08} and references therein. 

In this paper, we argue that it is inaccurate to calculate the gaseous absorption in both single- and mixed-gas atmospheres by combining individual gas cross-sections with large- and moderately-sized bins ($\Delta\nu = 25, 1$ cm$^{-1}$), or inefficient with high-resolution bins ($\Delta\nu = 0.01$ cm$^{-1}$). Throughout the paper we compare our calculations to the correlated-$k$ method, and verify our methods with a line-by-line calculation. We also consider the effects of including \ce{H2}-He pressure-broadening in the spectral calculations. 

Section one describes the methods and ingredients used in the correlated-$k$ tables (`$k$-tables') and cross-sections such as the pressure-broadening parameters and line lists used.

Section two verifies our methods with simple single-layer atmospheres and a line-by-line calculation. More realistic atmospheres are then used to contrast the spectra of the two methods (correlated-$k$ and cross-sections) and the effect of introducing \ce{H2}-He pressure broadening. The example cases include a simple Hot Jupiter, a typical late-T dwarf, and HD189733b in primary transit and secondary eclipse geometries.

Section three summarizes our findings.

\section{Methods}

\subsection{Line Lists}

A wide range of line list databases exists to provide the relevant molecular information for calculating spectra in the atmospheres of exoplanets and brown dwarfs. The key factors in deciding which line lists are most appropriate are the wavelength and temperature ranges for which they are valid. For example, a widely used line list database is the High Resolution Transmission (HITRAN) database \citep{rothman13}, which is collated from multiple experimental and theoretical sources. This database is, however, used mainly for representing the spectrum of the Earth, and is therefore only reliable for temperatures up to $\sim$400K, as it removes all insignificant line intensities at this temperature. Unfortunately, these insignificant line intensities become more significant with increasing temperature, meaning the relevant number of lines goes from thousands to billions as the temperature increases above 1500K. 

The ExoMol project \citep{tennyson12} contains a much more valid temperature (up to 1500-3000K) and wavelength range for the relevant gases involved in brown dwarf and exoplanet atmospheres. However, ExoMol line data are predicted solely from \textit{ab initio} calculations rather than measured in a laboratory and as such line positions and intensities may contain larger errors than those found experimentally in, e.g. HITRAN, for overlapping validity ranges. We note that HITEMP, a sister database to HITRAN, contain line lists appropriate for higher temperatures, and similarly contains a mixture of experimental and \textit{ab initio} data.

In Table~\ref{tbl:linelists}, we present a summary of the line lists that we selected according to the criteria above, where $Q(T)$ lists the chosen source of the partition function data, necessary to calculate the line intensities at the temperature of interest. We also use \ce{H2}-\ce{H2} and \ce{H2}-He collision-induced absorption from HITRAN 2012 \citep{richard2012} for any calculation containing \ce{H2}-He.

\subsection{Pressure-Broadening Coefficients}

Another measure of appropriateness for each line list is the validity of its pressure-broadening coefficients. Unfortunately, line list databases which are suitable for the pressure broadening found in \ce{H2}-He-dominated atmospheres are scarce. HITRAN's Earth-centric lists exhibit broadening parameters suited to air-broadening (\ce{N2} and \ce{O2}). ExoMol now provide \ce{H2} and He pressure broadening parameters for most of their gases. For all molecules ExoMol provides cross-sections with only Doppler (i.e. thermal) broadening, but no pressure broadening.

In order to correctly estimate the pressure broadening induced on spectral lines in brown dwarf and exoplanet atmospheres, we performed a literature search and found that \citet{amundsen14} had openly discussed their sources for \ce{H2}-He pressure broadening. This work was performed prior to ExoMol providing pressure broadening parameters for \ce{H2} and He, hence we have not used those values. Instead, most of the information found in Table~\ref{tbl:presbroad} that we use for our pressure broadening parameters is from \citet{amundsen14}, with a few additional sources added. Note that we have not chosen to implement the same procedure for Na and K, as they both produce massive broadening wings in the optical and are not experimentally well sampled for \ce{H2} and He broadening; instead we arbitrarily set the air-broadened widths to 0.075cm\textsuperscript{-1} atm$^{-1}$ based on experience with terrestrial radiative transfer studies. We also note that using Voigt lineshapes, as we have done for all gases, is especially dubious for the Na and K resonance lines \citet{burrows2001}. We intend on updating the Na and K $k$-tables in the future for more realistic lineshapes; however this will not drastically change the outcomes of this paper. For methods taken by other groups, see e.g. \cite{tremblin2015, baudino2015}.

In many cases these sources only provided broadening parameters for the lower rotational quantum number, $J_{low}$, up to 8-20 for the maximum  $J_{low}$ value depending on the gas, while our line lists contain data up to $J_{low}$ = 300. We implemented the broadening parameters into our line lists shown in Table~\ref{tbl:presbroad} by first converting them into a single `foreign broadening parameter' $\gamma_{0}$ assuming an atmospheric ratio of 85:15 for \ce{H2}:He, using the weighted sum $\gamma_{0} = \gamma_{H_2} \text{VMR}_{H_2} + \gamma_{He} \text{VMR}_{He}$, where VMR represents the volume mixing ratio. We then fitted this foreign broadening coefficient with a fourth order polynomial given by $\gamma_{0}(J_{low}) = \sum_{i=0}^{i=3} \alpha_i J_{low}^i$, where $\alpha_i$ represents each order's constant, up until the available data, then using the last available broadening coefficient for any $J_{low}$ higher than the maximum available. Note that \citet{amundsen14} used a linear approach up to the maximum $J_{low}$, and then a constant value as we have. 

\begin{figure}
    \centering
    \includegraphics[width=1.0\columnwidth]{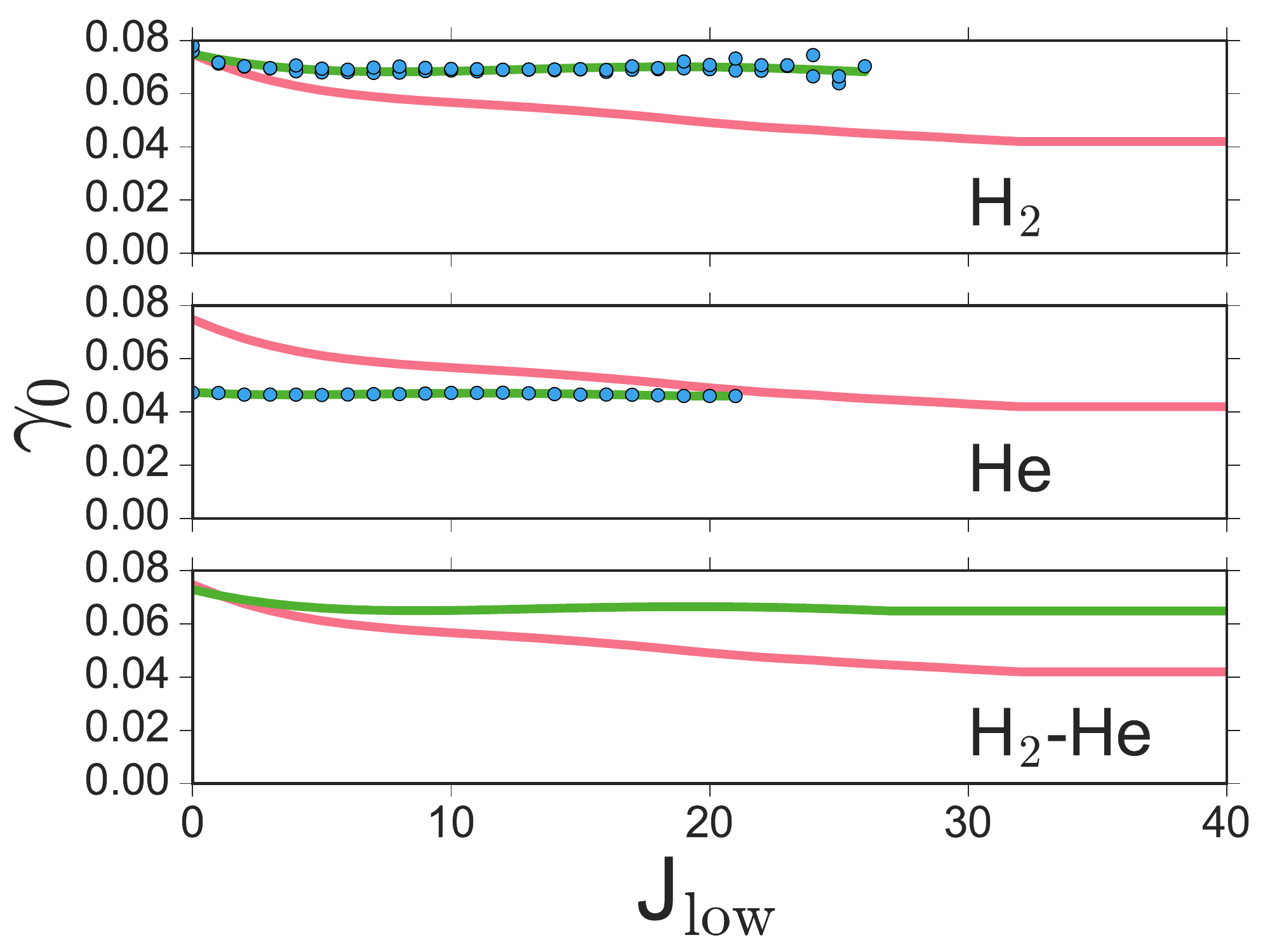}
    \caption{Pressure broadening coefficients for CO. Pink is air-broadening, blue circles are data points, green lines are fourth-order polynomial fits. Top panel: \ce{H2} broadening coefficients. Middle panel: He broadening coefficients. Bottom: \ce{H2}-He broadening coefficients with 85:15 ratio. }
    \label{fig:COpoly}
\end{figure}

This is of course not ideal, but does not introduce any complex error propagation which might be the case with a more sophisticated modelling that extrapolates to higher $J_{low}$. An empirical approach was considered, but while many gases show a gradual flattening of broadening coefficient with increasing $J_{low}$ \citep[see e.g.,][]{buldyreva2011collisional}, there is no clear or simple relationship that allows us to model all of the gases after this maximum. In general, the constant value appears to be a good first-order approximation. An example for the molecule CO is presented in Figure~\ref{fig:COpoly}, where we compare our new \ce{H2}-He foreign broadening to that provided for air by HITRAN. 

The pressure-broadened line half-width (cm\textsuperscript{-1}), is calculated from $\gamma = \gamma_0 \left(\frac{P}{P_0}\right) \left(\frac{T_0}{T}\right)^n
$ where the half-width at half-maximum of the Lorentzian profile $\gamma_0$ is determined at a standard temperature $T_0$ and pressure $P_0$ (i.e. 296K and 1 atm), $n$ is an empirically derived pressure-broadening temperature exponent found in Table~\ref{tbl:presbroad}, and $T$ and $P$ are the desired temperature and pressure of the line respectively. The temperature exponent is constant over all quantum rotational numbers.

\subsection{Cross-section Comparison}

In this subsection we investigate two methods of solving the radiative transfer equation regarding optical depth, and show how they affect the spectra produced by forward models of brown dwarfs and exoplanets. As our retrievals are entirely dependent on our forward models being `correct', this is in an important point to address.

We now describe the exact details of the $k$-tables we use in our radiative transfer calculations.

For each gas in Table~\ref{tbl:linelists}, we must calculate $k$-tables to be used in the correlated-$k$ method of solving the radiative transfer equations. We developed a method to remove insignificant lines in the line databases at specified temperatures, as this will shorten the length of computation time of the $k$-tables. The method first calculates the line intensities at a particular temperature, and orders the line intensities from smallest to largest. It then creates a cumulative sum from the smallest to largest line intensities, and removes the smallest n\% of contributions to the total line intensities. We use n = \num{1e-10}, a very small percentage, so that we do not underestimate any continuum effects. We do this for our entire temperature range, i.e. 100 - 2950K, over 20 equally spaced (150K) temperature points as different lines become important at different temperatures. At the lower temperatures, a majority of the lines are stripped away (leaving $\sim$ \num{1e4} lines), while at higher temperatures the line lists are essentially identical (up to $\sim$ \num{1e11}).

For the $k$-tables, we first calculate the underlying absorption spectrum monochromatically (at least 1/6 Voigt line width) over the entire spectral region (0.3-30\textmu m). We calculate the absorption and hence cumulative $k$-distributions at output wavelengths having separation $\Delta \lambda$=0.001\textmu m, with square bins of width double that of the separation (i.e. a resolution of $\lambda$=0.002\textmu m). We use Gauss-Lobatto quadrature to sample the distributions. To include the contribution of wings of lines centred outside the spectral region of interest, the total wavelength range considered is defined as a range between $\nu_{min} - \nu_{cut}$ to $\nu_{max} + \nu_{cut}$. The total spectral interval is then subdivided into 1.5cm$^{-1}$ bins where line data is stored. Absorption at a particular output wavelength is then calculated by considering lines stored in the adjacent bins and the bin in the middle. Wing contributions from the lines centred outside these bins are calculated at the middle and end bins using a quadratic polynomial (because the wing shape follows a Lorentzian), and added on to the absorption at the output wavelength. For each line calculation, a line wing outside the cutoff (25cm$^{-1}$ from centre) is ignored.

We calculate a grid of spectral opacities with the 20 aforementioned temperature points, and 20 pressure levels equally spaced in logspace from $\sim$ \num{1e-7} to 100 atm, using 20 g-ordinates for Gauss-Lobatto quadrature. The lineshapes are given by a Voigt profile, where the line-wing cut-off ($\nu_{cut}$) is at 25cm$^{-1}$ for all gases but alkali, where the cut-off is at 6000cm$^{-1}$.

The pressure-temperature grid is linearly interpolated in temperature-log-pressure space to find the desired spectral opacity. The $k$-tables may also be resampled into lower resolutions to increase calculation speed, if necessary. For a more detailed description of how these calculations are done, the reader is referred to \citet{nemesis08}. These $k$-tables calculated here will be adopted in future works.

For the cross-section calculations in the literature, ExoMol \citep{hill2013} calculate a high-resolution (<0.01cm$^{-1}$) cross-section and the integral of the cross-section is preserved for all requested resolutions. \cite{macdonald2017} calculate their cross-sections at 0.01cm$^{-1}$ (from \cite{hedges2016}), and bin them down (presumably preserving area) to 1cm$^{-1}$ resolution before using them. \cite{line2016b} use a 1cm$^{-1}$ resolution opacity sampling method on pre-computed cross sections which have a variable resolution wavenumber grid that samples the lines at 1/4 of their Voigt half widths from \cite{freedman08, freedman2014}.

\begin{figure}
 \centering
 \includegraphics[width=1.0\columnwidth]{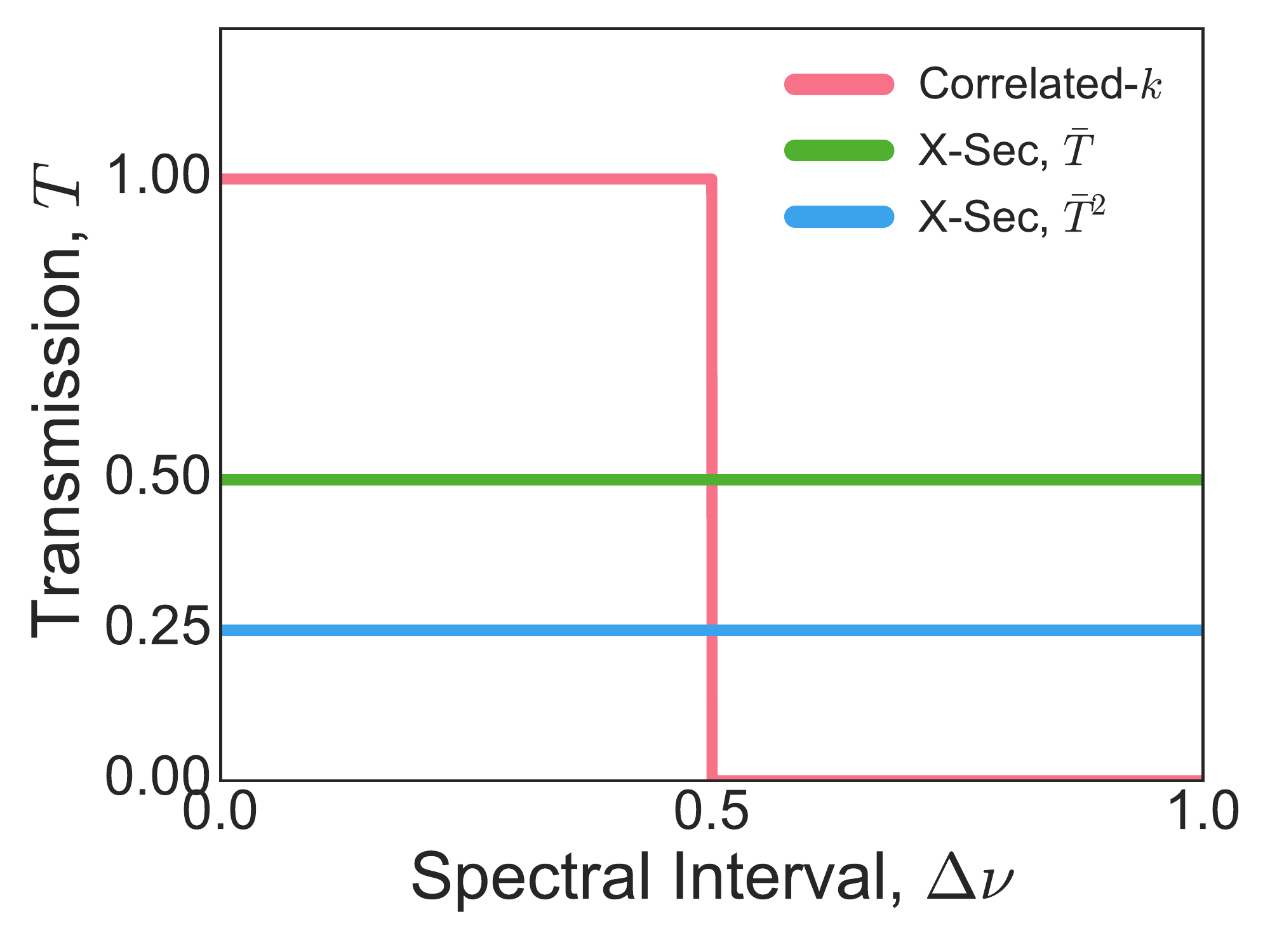}
 \caption{An illustrative example of the effects on transmission when multiplying monochromatic and non-monochromatic wavelength regions. The Correlated-$k$ transmission (pink) squared is identical to the non-squared, as the first half of the interval is 1 squared, and the second half is 0 squared. The cross-section transmission (green) is averaged over the bin, such that when squared it becomes a yet smaller value (blue).}
 \label{fig:trans}
\end{figure}

One of the major issues with low- and medium-resolution cross-sections is that they cannot combine gases effectively due to their insufficient resolution. The usual approach is to sum the individual contributions of the various gases and weight them by their volume mixing ratio \citep{sharp07, waldmann15}. The multiplication property of transmission is only valid when these calculations are done monochromatically. To illustrate this issue for low- and medium-resolution cross-sections, consider the simple two-gas problem in a single layer where each gas has an identical transmission spectrum over a given interval which is larger than the bin size for the cross-section, $\Delta \nu$, shown in Figure~\ref{fig:trans}. In this extreme example, the multiplication using these two methods produces mean transmissions that are a factor of two different. Correlated-$k$ preserves this multiplication property of transmissions \citep{goody1989atmospheric}, i.e. the calculations are the same as for monochromaticity. Using cross-sections invokes using a mean transmission over the interval and hence the multiplication can produce fallacious results. 

The opacity sampling method requires a sufficient number of samples within a specified bin to correctly estimate the area of the cross-section, though this number is ill-defined and depends on spectral resolution. \cite{line2016b} show that the 1cm$^{-1}$ resolution opacity sampling is sufficient for their purposes (1.0-2.5\textmu m); however, this may not be a good resolution for either higher spectral resolutions or longer wavelengths. We briefly explore the effects of using opacity sampling as a means of computing the gaseous absorption in the next section, but do not include it in our main results section as it would require a large overhaul of our code.

There are two key issues with using the Doppler-broadened cross-sections provided by the ExoMol project \citep{hill2013}. One is that they fundamentally should not be used to solve the radiative transfer equation in mixed-gas atmospheres by combining single gas cross-sections as they are either ineffective (low- and medium-resolution) or computationally expensive (high-resolution). The second is that pressure-broadening is a key property in calculating atmospheric spectra that should not be ignored. These points will be addressed in the next section.

\begin{figure}
 \centering
 \includegraphics[width=1.0\columnwidth]{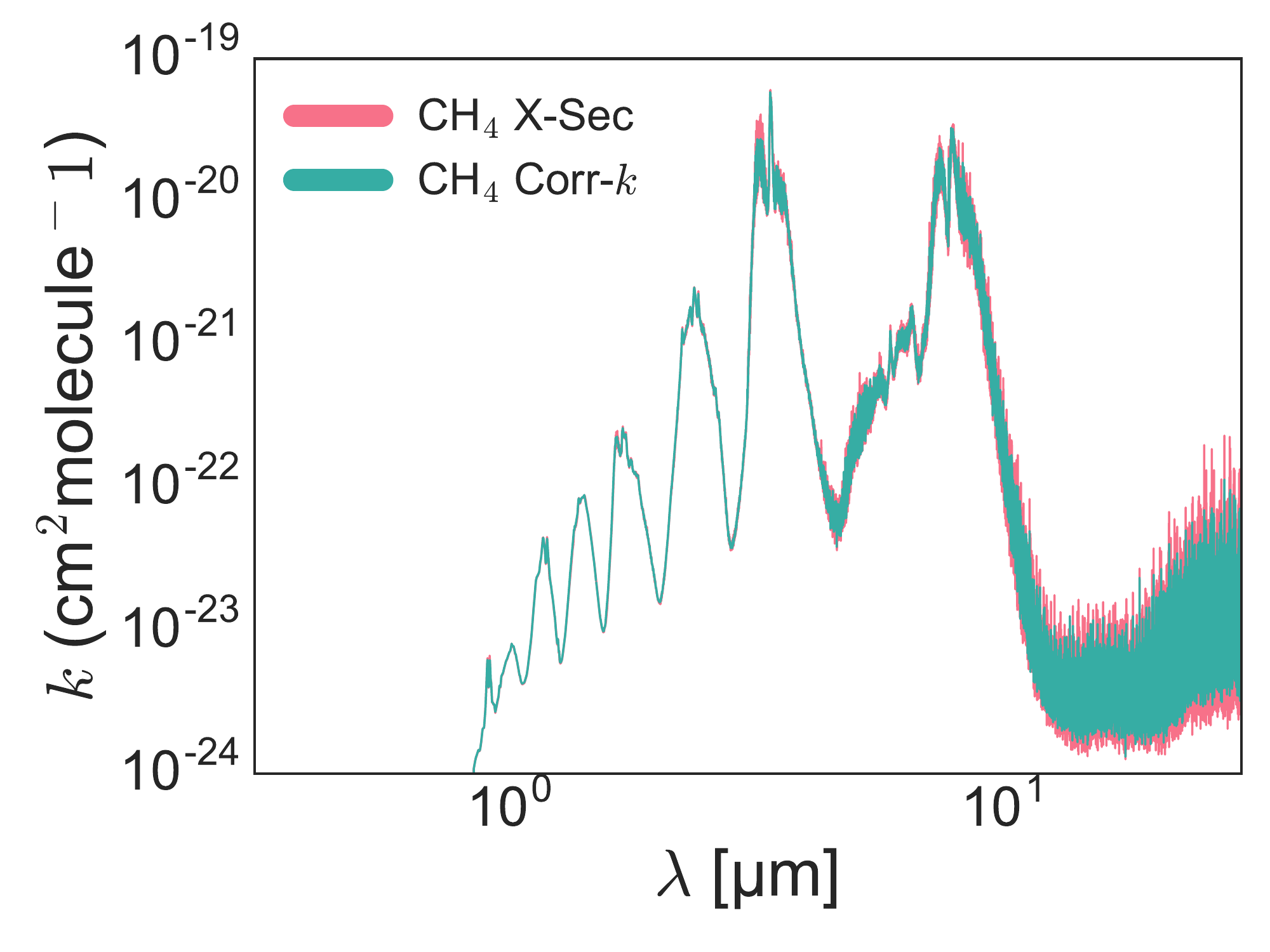}
 \caption{The \ce{CH4} $k$-table opacity at low-pressure (blue) is compared to the Doppler-broadened cross-sections at a resolution of $\Delta \nu = 1$cm$^{-1}$ from the ExoMol project (pink) for 1600K. The spectra are binned to $\Delta \lambda = $0.005\textmu m resolution.}
 \label{fig:exmkch4-1}
\end{figure}

At low pressures, where all of the pressure-broadening has ceased and the lineshape is dominated by Doppler broadening, the mean opacity (i.e. $\bar{k} = \sum_{i=1}^{NG} k_i \Delta g_i $) of our $k$-tables should be equivalent to the ExoMol cross-sections. We show this to be true for \ce{CH4} at 1600K in Figure~\ref{fig:exmkch4-1}, where the cross-sections were taken from www.exomol.com at a resolution $\Delta \nu = 1$cm$^{-1}$. The spectra are binned to $\Delta \lambda = $0.005\textmu m resolution. This is also the case for the other relevant gases.

By flattening our $k$-distribution, i.e. replacing $k_i$ for each g-ordinate with $\bar{k}$ so that the $k$-distribution $k(g)$ is now a flat (constant) distribution, our $k$-tables at these low pressures are mathematically identical to using cross-sections, and so will fall foul to the aforementioned gas-mixing issues.

Instead of using the ExoMol cross-sections directly where we are limited to the temperatures ranges supplied online, we carry out our comparison using only the lowest pressure points of our flattened $k$-tables as a proxy. In the next section we show first that this is a good approximation to using cross-sections of resolution $\Delta \nu = 1$cm$^{-1}$.

\section{Results}

\begin{figure}
 \centering
 \includegraphics[width=1.0\columnwidth]{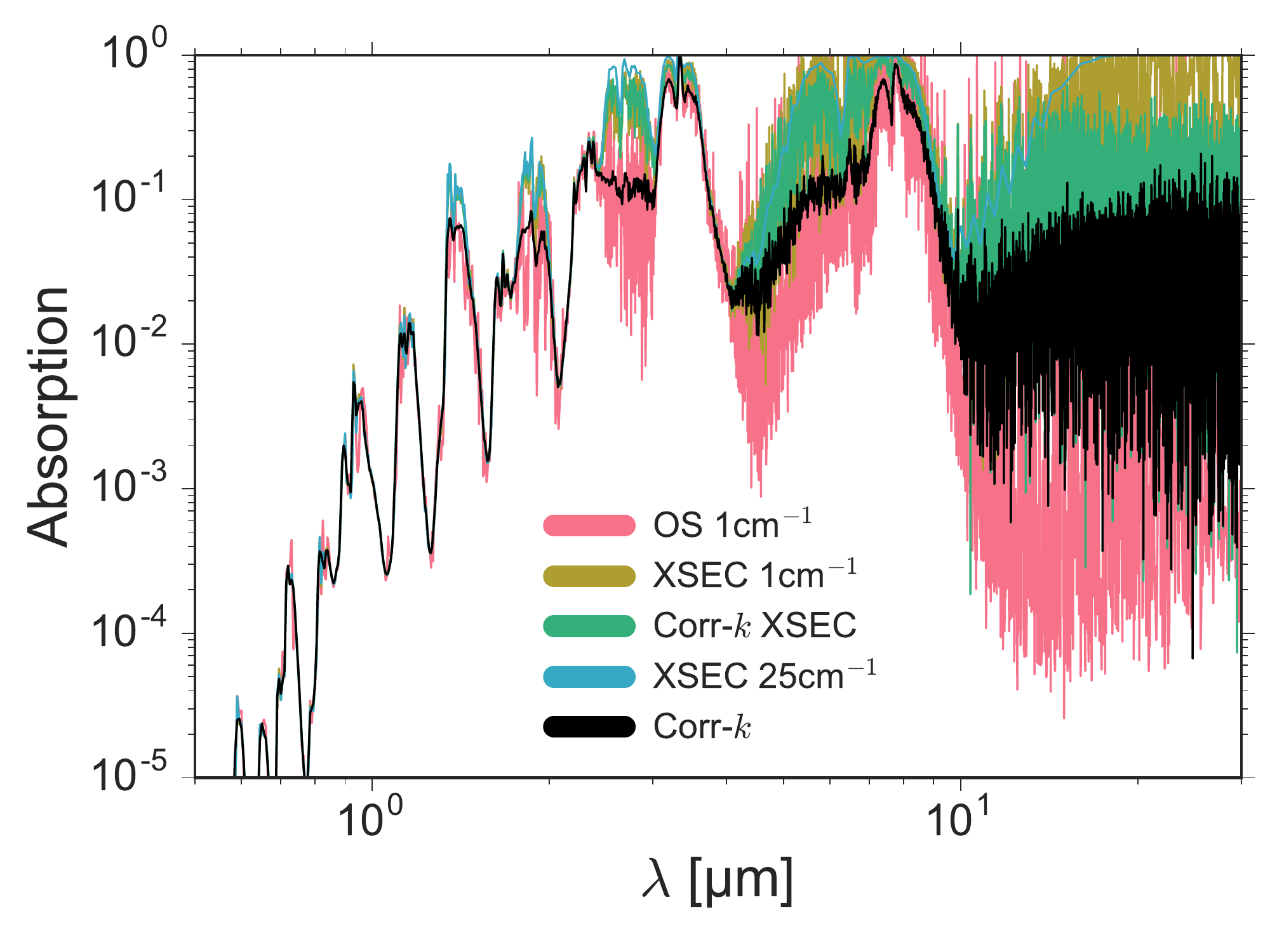}
 \caption{Absorption calculated for a layer consisting of 50\% \ce{CH4}, 50\% \ce{H2O} at 1000K and path amount of \num{1e20} molecule cm$^{-2}$. Pink is opacity-sampling at 1cm$^{-1}$ resolution using the 0.01cm$^{-1}$ resolution ExoMol cross-sections. Gold and blue are the cross-section absorption spectra at 1cm$^{-1}$ and 25cm$^{-1}$ resolution respectively, black line is the correlated-$k$ method with a resolution of 0.002\textmu m the green line is the cross-section derived from our correlated-$k$ method.}
 \label{fig:abs}
\end{figure}

First we would like to show the inability to effectively mix gases with cross-sections. We create a simple one-layer model at low pressure (and therefore only Doppler-broadened), with a temperature of 1000K, and containing 50\% \ce{H2O} and 50\% \ce{CH4}. We calculate the mean transmission for the correlated-$k$ using:

\begin{equation}\label{eqn:trank}
\bar{T} = \sum_{i=1}^{NG} \sum_{j=1}^{NG} e^{- (k_i m_a + k_j m_b) } \Delta g_i \Delta g_j
\end{equation}

and the cross-sections:

\begin{equation}\label{eqn:tranxsec}
\bar{T} =  e^{- (k_a m_a + k_b m_b) } 
\end{equation}

where $i$ and $j$ represent the index of the weights and $a$ and $b$ are labels for the first and second gas respectively, $k$ is the absorption coefficient (cm$^{2}$ molecule$^{-1}$), and $m$ is the absorber amount (molecule cm$^{-2}$). Absorption (1 - $\bar{T}$) is calculated with four different methods: 1) using our $k$-tables described earlier, the online ExoMol cross-sections directly at two different resolutions - 2) 1cm$^{-1}$ and 3) 25cm$^{-1}$); 4) our cross-section proxy $k$-tables described in the previous section; and 5) the opacity-sampling method at 1cm$^{-1}$ resolution using the 0.01cm$^{-1}$ ExoMol cross-sections. Methods 1 and 4 are combined using equation~\ref{eqn:trank}, while methods 2,3 and 5 are combined using equation~\ref{eqn:tranxsec}. The spectra are then binned to $\Delta \lambda = $0.005\textmu m resolution, and shown in Figure~\ref{fig:abs}. We choose these two resolutions for the cross-sections because the prior is the usual resolution used by other modellers, and the latter because it is closer to the typical resolution that we are presenting in Figure~\ref{fig:mixedgases} and subsequent figures in the visible ($\sim$0.6\textmu m). Beyond $\sim$3\textmu m, our $k$-tables are in fact higher resolution than $\Delta \nu = 1$cm$^{-1}$. This means that we expect the cross-sections (and opacity sampling method) to become relatively less accurate at longer wavelengths. This is also generally true because the Doppler width is proportional to the wavenumber, i.e. at long wavelengths lines become increasingly narrow and consequently require a higher resolution to be resolved.

\begin{figure}
 \centering
 \includegraphics[width=1.0\columnwidth]{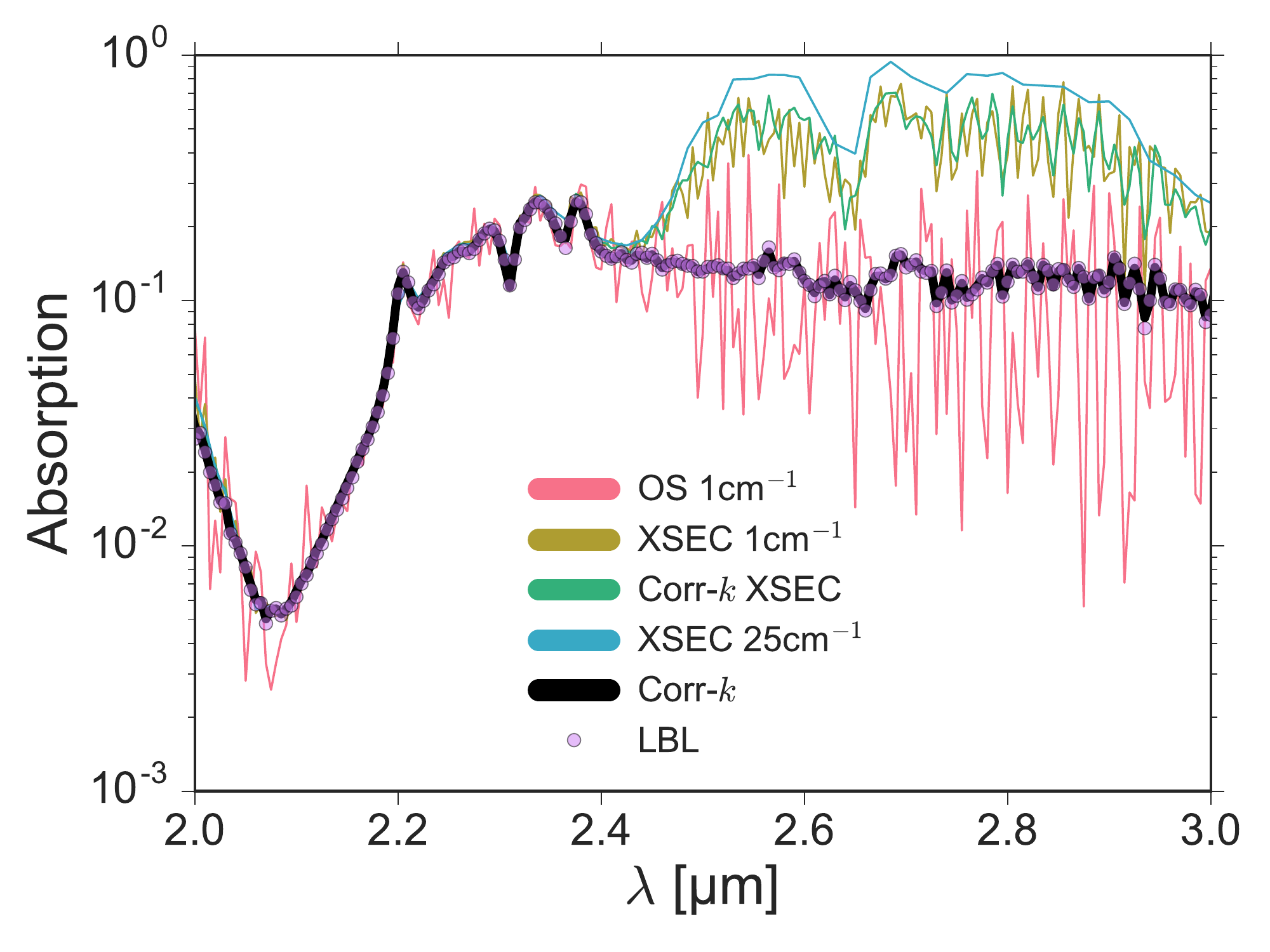}
 \caption{Same as Figure~\ref{fig:abs} but zoomed in to 2-3\textmu m. The purple dots are a line-by-line calculation calculated at $\Delta \nu = 0.01$cm$^{-1}$.}
 \label{fig:abszoom}
\end{figure}

\begin{figure}
 \centering
 \includegraphics[width=1.0\columnwidth]{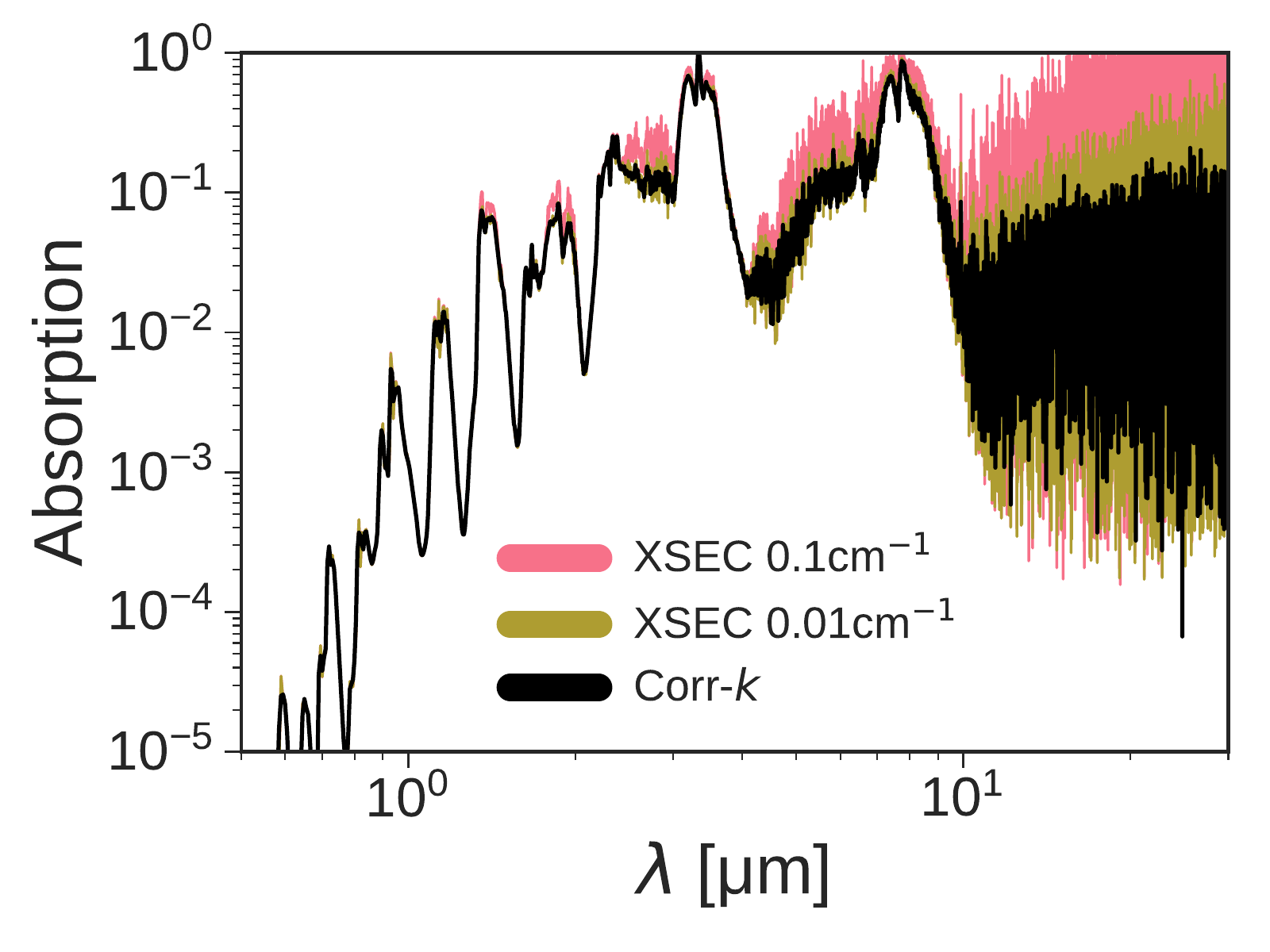}
 \caption{Similar to Figure~\ref{fig:abs}, except pink is the cross-section absorption spectra at 0.1cm$^{-1}$ resolution, gold is cross-section at 0.01cm$^{-1}$ resolution, and black is the same correlated-$k$ calculation}.
 \label{fig:abshigh}
\end{figure}

In Figures~\ref{fig:abs} and~\ref{fig:abszoom} we see that the opacity sampling method produces absorption that is approximately of the same order as the correlated-$k$ and line-by-line methods. However it also exhibits a slight underestimation of the absorption and a large increase in noise as we go to longer, less-sampled wavelengths. The effect is especially pronounced in certain bands such as the 2.5-3\textmu m band in Figure~\ref{fig:abszoom}. While these effects are interesting, it would require a major overhaul of our code to further investigate, and hence do not include it in further analysis. Figure~\ref{fig:abs} also shows that the overlapping bands can cause up to $\sim$2 orders of magnitude change in absorption when comparing the correlated-$k$ and cross-section methods. We believe these `ghost features' are a consequence of one-cross section being more accurate than the other, so that in a mixture, when the inaccuracy dominates the total absorption, a ghost feature is formed. This is because the average transmissions computed from equation~\ref{eqn:tranxsec} does not capture the non-linear relationship between transmission and opacity. As expected, beyond 10\textmu m the absorption varies enormously due to an increase in the number of binned lines per resolution element in the cross-sections. Also it shows that our $k$-table proxy cross-sections show approximately the same (if not smaller) effects than the two different ExoMol cross-sections, hence we can conclude that using these cross-sections are a reasonable proxy to use for the rest of this paper. As shown in Figure~\ref{fig:abszoom}, the line-by-line calculation between 2-3\textmu m agrees very well with our correlated-$k$ method and shows that there is a real concern for combining cross-sections at a resolution of $\Delta \nu = 1$cm$^{-1}$, especially at longer wavelengths. By comparing our correlated-$k$ method to a high-resolution cross-section (0.01cm$^{-1}$) in Figure~\ref{fig:abshigh}, it is evident that they now share a similar spectral morphology; however, the increase in resolution of the cross-section now causes a large increase in computation time. Figure~\ref{fig:abshigh} also shows a medium-resolution case (0.1cm$^{-1}$) which shows inaccuracies beginning to occur in a non-negligible way. In a vast majority of real-life data resolutions, using the correlated-$k$ method is faster or more accurate that using these high- and medium-resolution cross-sections because the the correlated-$k$ method can be precalculated at exactly the right resolution of the data, whereas the cross-section method must always be at a much higher resolution than that of the data.

\subsection{Primary Transit}

In this example we take a simplified but more realistic atmosphere found in exoplanetary science, and compare its resulting spectra using the two different gaseous absorption methods.
We do this using an isothermal (2000K) Hot Jupiter-esque primary transit as our example ($1.06M_{Sat}$, $1.4 R_{Jup}$ around a Solar-size star), which we refer to as PT1. 

Here we limit the $k$-tables to have only Doppler broadening by using the lowest pressure point in our grid (\num{1e-7}atm). This means we are directly comparing $k$-tables with cross-sections, i.e. the effect of flattening over the $g$-distribution. From this section onwards, we also include collision-induced absorption from \ce{H2}-\ce{H2} and \ce{H2}-He in our realistic atmosphere calculations where appropriate.

\begin{figure}
 \centering
 \includegraphics[width=1.0\columnwidth]{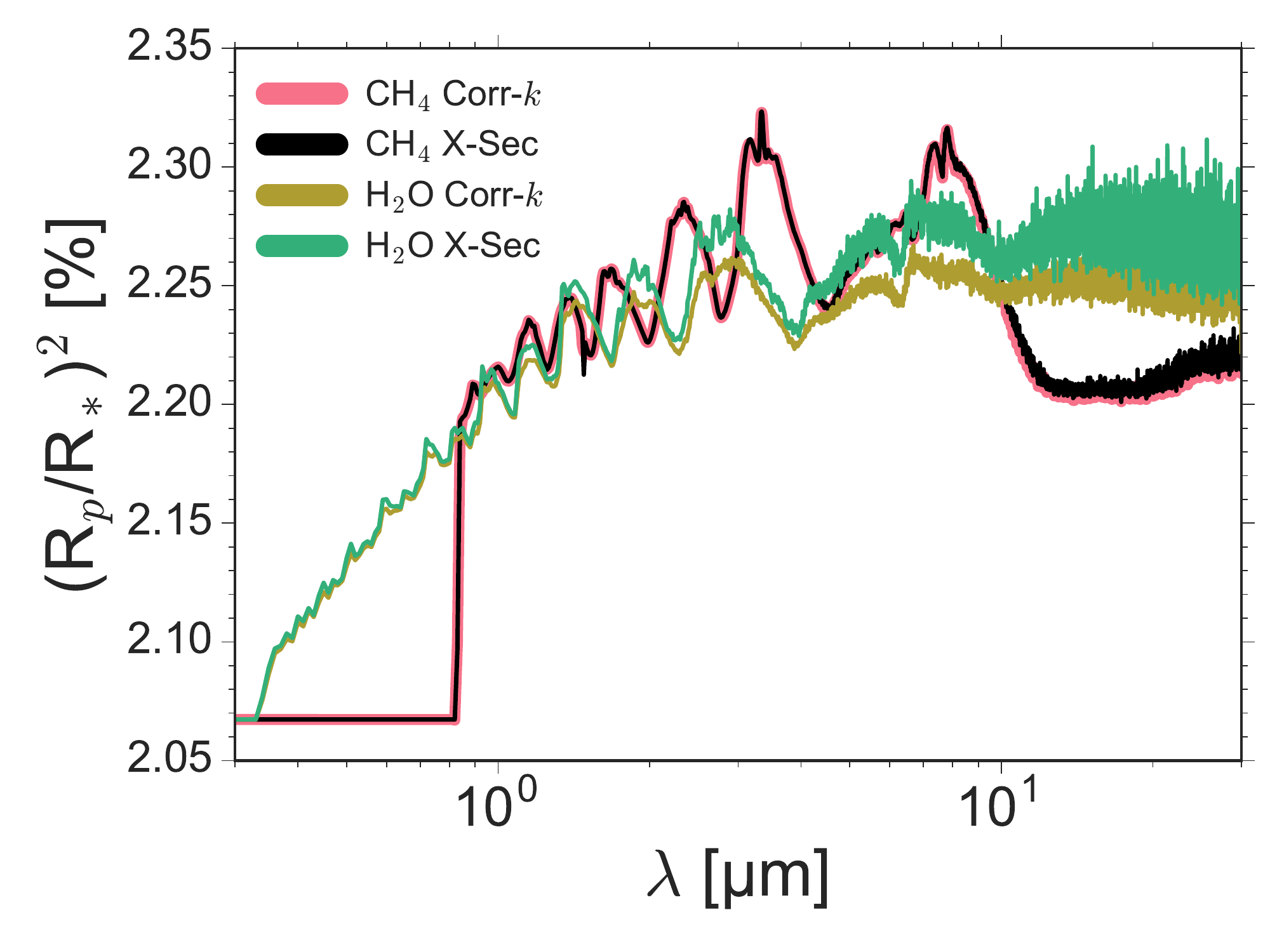}
 \caption{Spectra for two PT1 (defined in manuscript) exoplanets consisting of a single gas (100\% \ce{CH4} in pink and black, 100\% \ce{H2O} in gold and green) calculated using two methods. Legend labels X-Sec and Corr-$k$ represent cross-sections and correlated-$k$ (Doppler-broadened only).}
 \label{fig:singlegas}
\end{figure}

In Figure~\ref{fig:singlegas}, we have two different atmospheres, each with a set of spectra calculated by the correlated-$k$ method and the cross-section method. We have a 100\% \ce{CH4} atmosphere, and a 100\% \ce{H2O} atmosphere. Here, no mixing of the gases is required, and therefore the two different methods produce almost identical results, with the differences being more pronounced for \ce{H2O}. These results are relatively similar because the propagating error in transmission multiplication throughout the pressure-varying path due to bin-averaging is small compared to those when averaging with overlapping gaseous bands. Note that the changes are not due to pressure-broadening effects, as the $k$-tables only contain Doppler broadening. 

\begin{figure}
 \centering
 \includegraphics[width=1.0\columnwidth]{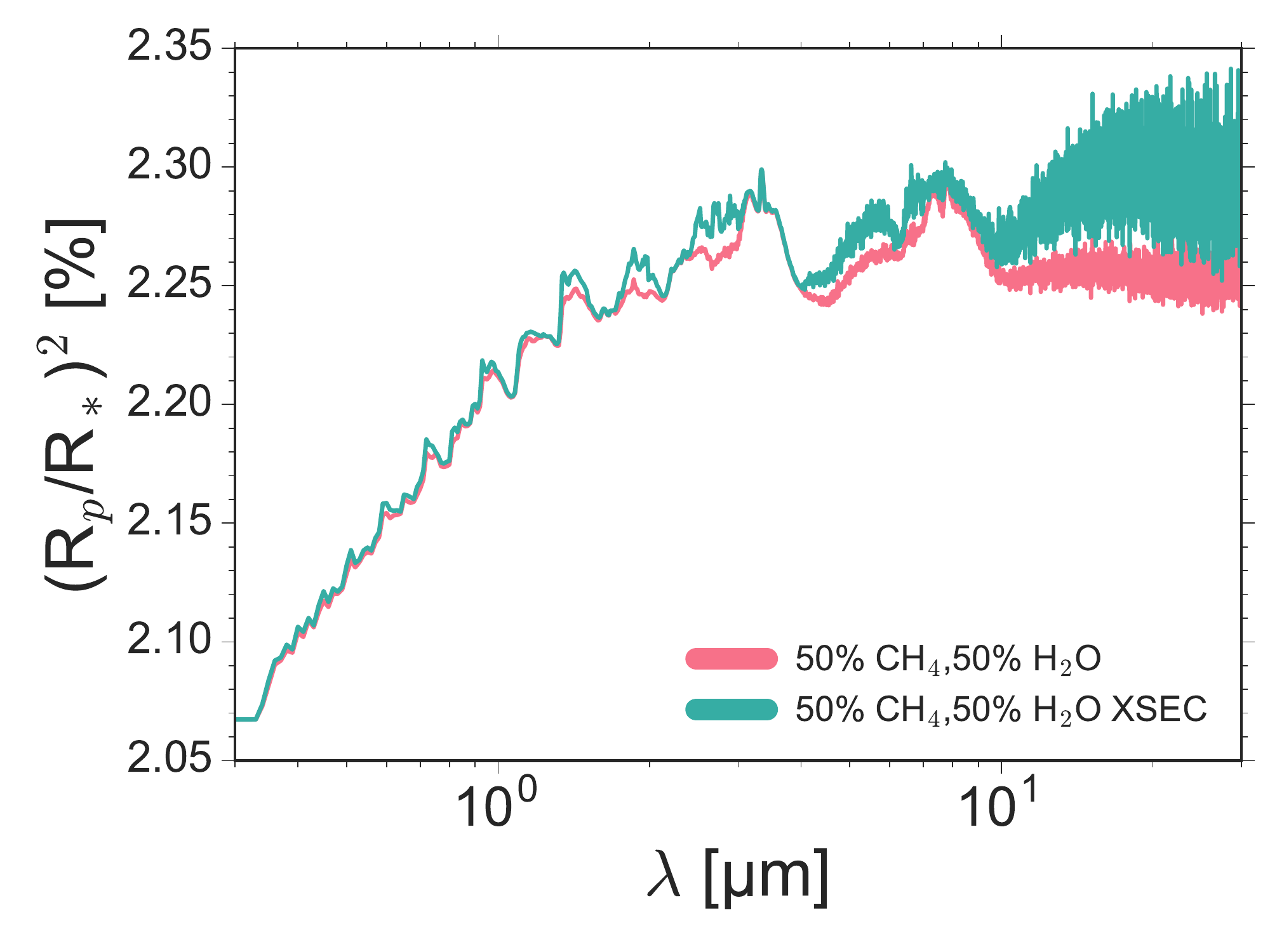}
 \caption{Spectra for one PT1 exoplanet consisting of 50\% \ce{CH4}, 50\% \ce{H2O}. Pink line is the correlated-$k$ method (Doppler-broadened only), blue is the cross-section method.}
 \label{fig:mixedgases}
\end{figure}

We introduce a third atmosphere, composed of 50\% \ce{CH4} and 50\% \ce{H2O}. In Figure~\ref{fig:mixedgases}, we compare the spectra produced by the Doppler-broadened correlated-$k$ and the cross-section methods. We find that not only do the cross-sections overestimate the transit depth, but indeed they change the morphology of the spectrum itself, similar to the results for in Figure~\ref{fig:abs}.

\subsection{Brown Dwarfs}

In the previous two subsections we verified our correlated-$k$ method via line-by-line calculations, and showed the impact of various resolutions of cross-sections on the morphology of the spectra and absorption profiles of simple exoplanet atmospheres. In this subsection and the next we intend to show two things for multiple types of realistic \ce{H2}-He-dominated atmospheres: 1) the effects of using cross-sections to incorrectly mix gases, and 2) the effects of including pressure-broadening on the resulting spectra. 

\begin{figure}
 \centering
 \includegraphics[width=1.0\columnwidth]{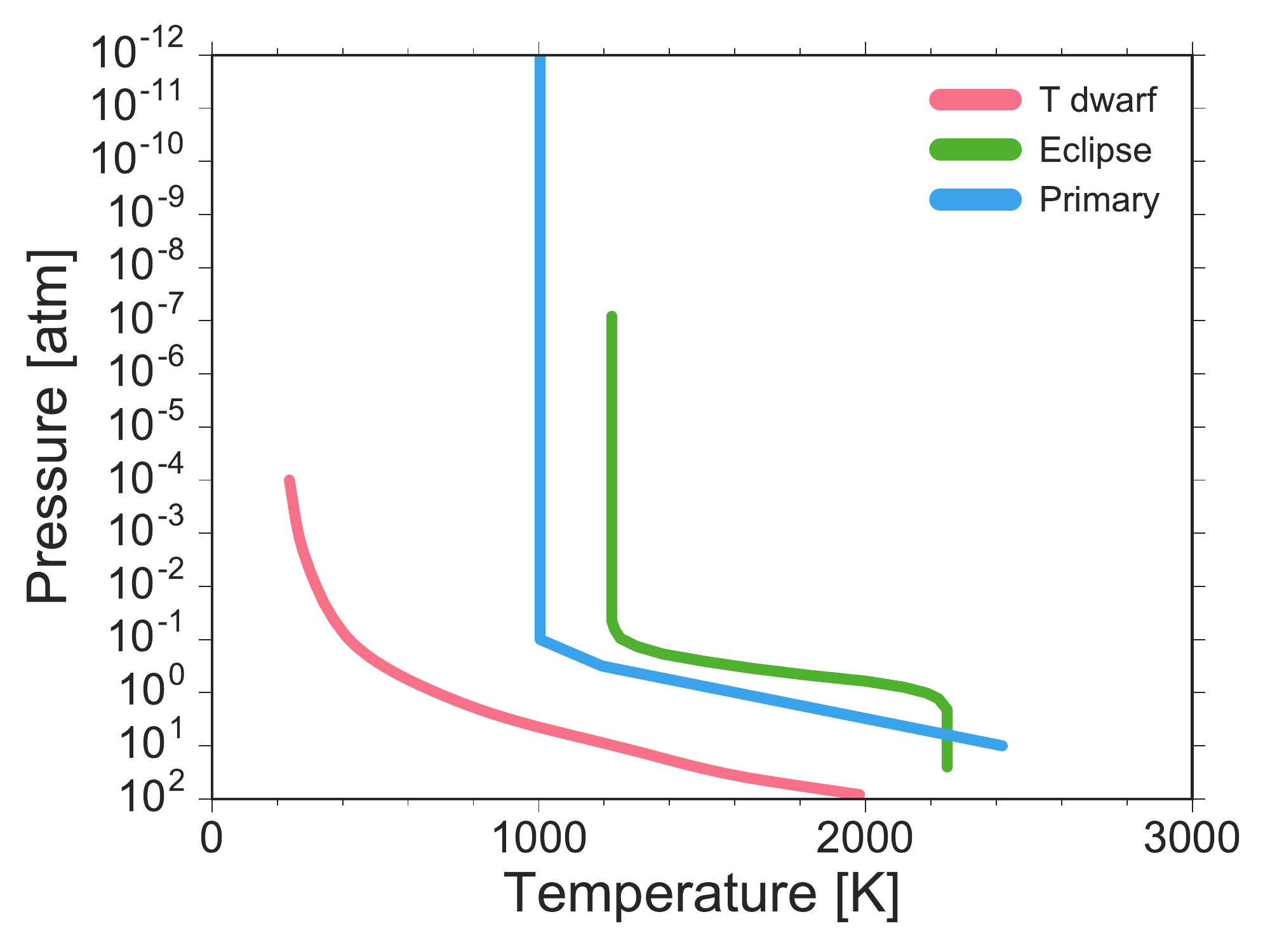}
 \caption{Temperature profiles for the brown dwarf (`T dwarf' in pink), the primary transit of HD189733b (`Primary' in blue) and secondary eclipse of HD189733b (`Eclipse' in green).}
 \label{fig:tpprofs}
\end{figure}

To illustrate the effects of using cross-sections, correlated-$k$ with no pressure-broadening (smallest pressure level available at our $k$-tables, $\sim$1e-7 atm), and correlated-$k$ with pressure-broadening to calculate the spectra of brown dwarf atmospheres, we use a typical late-T dwarf as an example, as it contains significant amounts of \ce{H2O}, \ce{CH4}, and \ce{NH3} (\num{3.5e-4}, \num{4.0e-4}, \num{2.3e-4} respectively for their volume mixing ratios, which are representative values that are constant in height). The spectrum also contains Na and K with VMRs of \num{3.5e-5} and \num{3.5e-6}. The mass and radius are 41.5M$_{Jup}$ and 1.38R$_{Jup}$ respectively, and the temperature profile is a 673.5K grid model from \citet{saumon08} with the appropriate surface gravity. The temperature profile, along with those from the next two sections, is shown in Figure~\ref{fig:tpprofs}. The exact numbers for these parameters do not significantly change the outcome of the results, and are the best fit parameters of a previous unpublished retrieval performed on the object GL570D. 

\begin{figure}
 \centering
 \includegraphics[width=1.0\columnwidth]{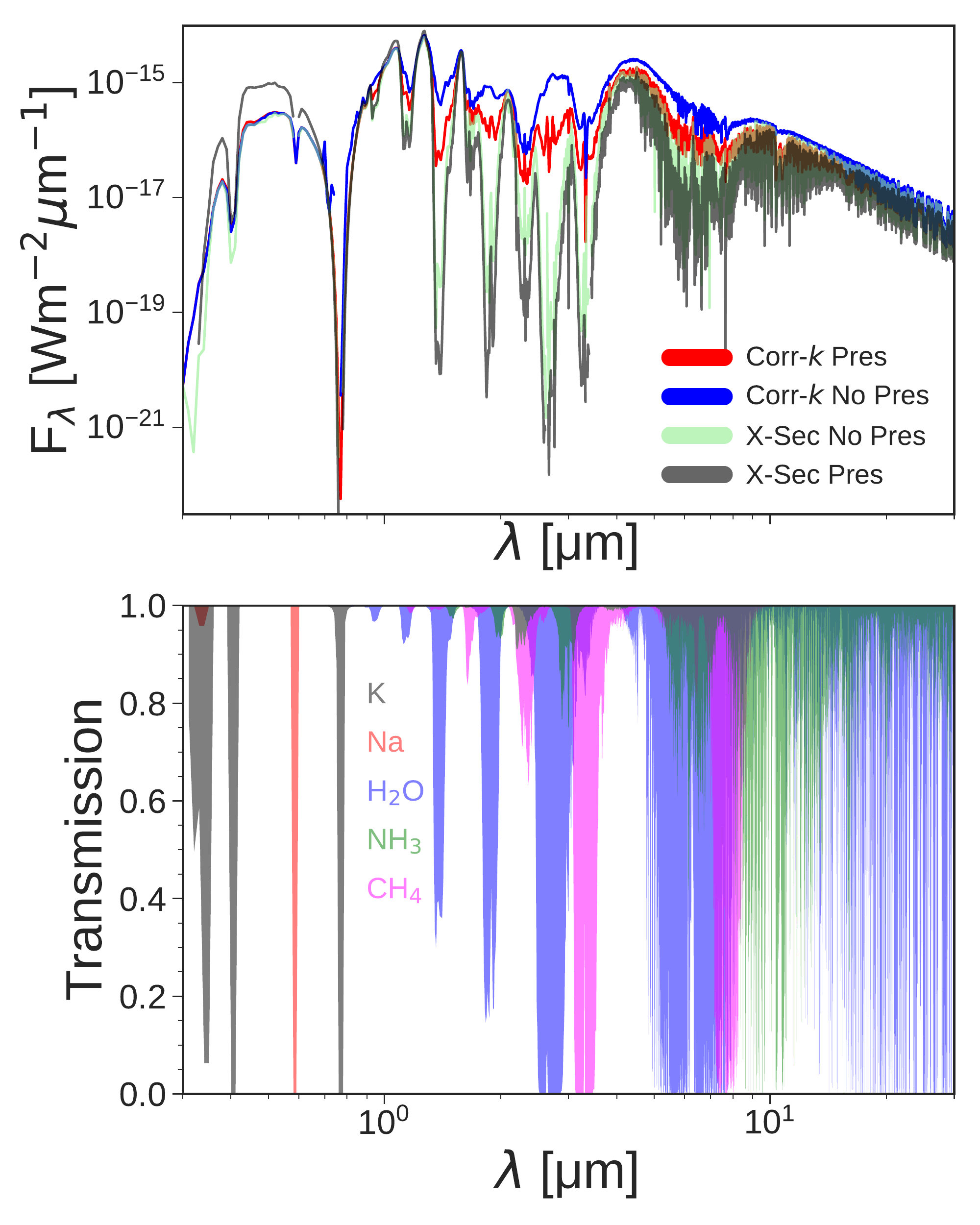}
 \caption{Top plot: Spectra calculated from best fit parameters for GL570D using various methods of calculation. The spectra are calculated for cross-sections with (translucent green) and without pressure-broadening (translucent black), and correlated-$k$ with (red) and without pressure broadening (blue). Bottom plot: the transmission calculated for the three major absorbing species in the atmosphere in a single layer of atmosphere at 700K, where each gas has been weighted by its volume mixing ratio for a given path amount.}
 \label{fig:TD}
\end{figure}

Figure~\ref{fig:TD} contains two plots. In the upper plot, we have spectra calculated using the cross-section method with and without pressure broadening, and the correlated-$k$ method with and without pressure broadening. In the bottom plot, we have calculated the transmission in a single layer at 700K and 1 atm, weighted by the volume mixing ratios with a given path amount (\num{1e23} molecules/cm$^{-2}$) for \ce{NH3}, \ce{CH4}, \ce{H2O}, Na and K, so that it can be easily seen which spectral features belong to which species.

In all of the large bands we find a discrepancy between the cross-section method and the correlated-$k$ method, usually of multiple orders of magnitude. For example, the 2.8\textmu m water band produces luminosities that vary by up to $\sim$3-4 orders of magnitude. The same can be said for the 3.2\textmu m \ce{CH4} band. In molecular bands, the addition of pressure broadening can produce an additional order of magnitude change. These are huge effects that would certainly be reflected in parameter estimation during retrievals. 

\subsection{Primary Transit and Secondary Eclipse}

\begin{figure}
 \centering
 \includegraphics[width=1.0\columnwidth]{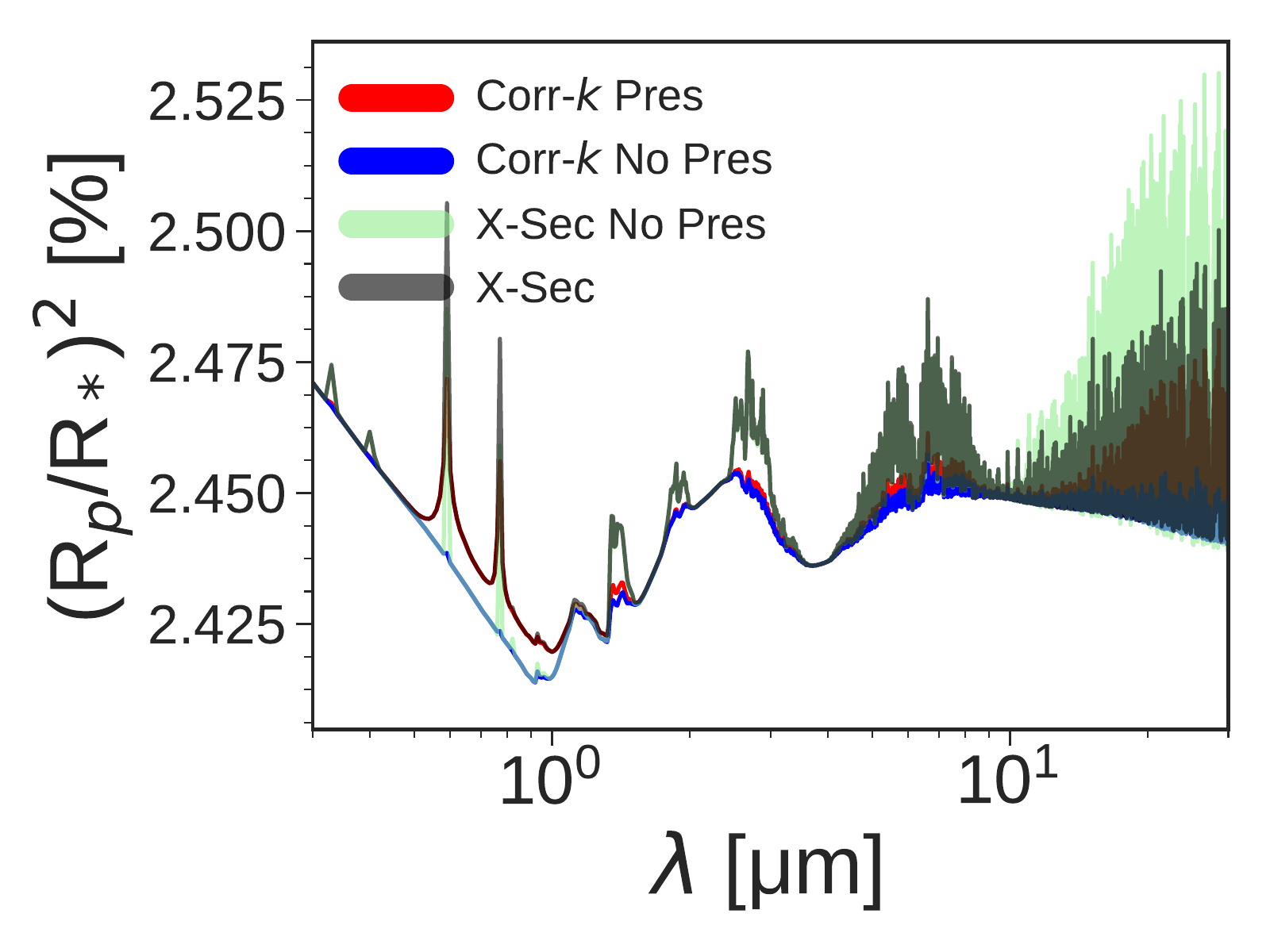}
 \caption{A fiducial HD189733b atmosphere for primary transit. The spectra are calculated for cross-sections with (translucent green) and without pressure-broadening (translucent black), and correlated-$k$ with (red) and without pressure broadening (blue).}
 \label{fig:PTHD}
\end{figure}

In Figure~\ref{fig:PTHD} we present a fiducial model from \citet{barstow2014} of the primary transit spectrum of HD189733b, calculated using the three methods as described before for the brown dwarfs. The atmosphere primarily consists of the usual \ce{H2}, He, \ce{H2O}, Na and K gases, along with a haze to cover the observed Rayleigh slope. 

Similar to the brown dwarf cases, we find that the total opacity and morphology of the spectra differ greatly between the methods. The transit depth can vary by up to 1\%, as seen at the 2.8\textmu m water band. Contrary to the brown dwarfs, we note that the pressure broadening effects are subdued for primary transit. This is because we are probing much higher in the atmosphere than in the brown dwarf case, where Doppler broadening is the main agent of broadening. The Na and K lines are changed vastly, although we must note that the visible region is especially subject to large changes in $\Delta \nu$ (a constant $\Delta \lambda$, as we have, produces larger $\Delta \nu$ in the visible), and thus the effects are more pronounced than for the infrared. 

\begin{figure}
 \centering
 \includegraphics[width=1.0\columnwidth]{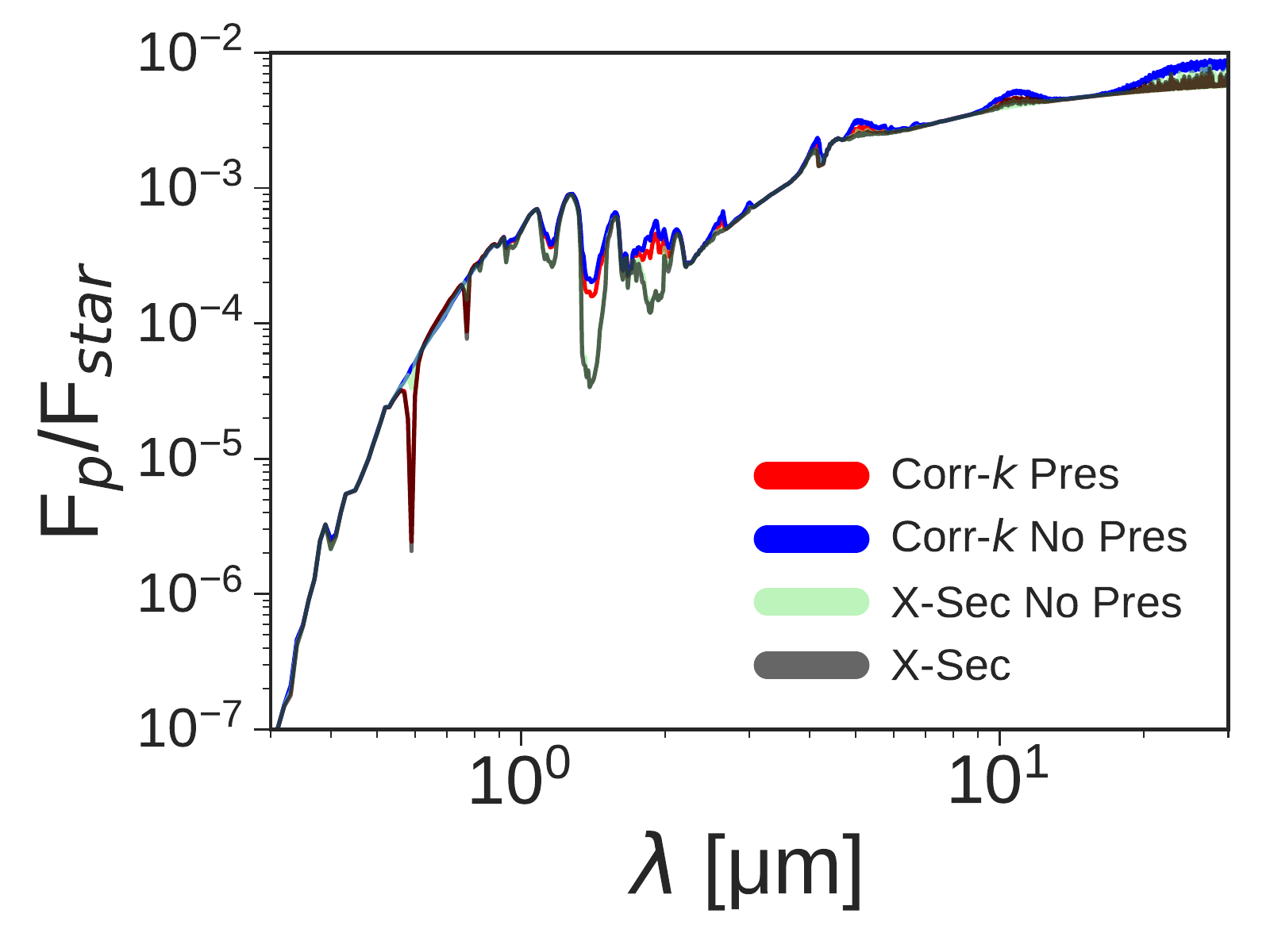}
 \caption{A fiducial HD189733b atmosphere for secondary eclipse. The spectra are calculated for cross-sections with (translucent green) and without pressure-broadening (translucent black), and correlated-$k$ with (red) and without pressure broadening (blue).}
 \label{fig:SECHD}
\end{figure}

For secondary eclipse, the VMRs used are slightly different than for primary transit. \ce{H2O}, \ce{CO2}, \ce{CO} and \ce{CH4} all have VMRs of \num{1e-4}. \ce{H2}, He, Na, and K have VMRs of 0.9, 0.1, \num{5e-6}, and \num{1e-7} respectively. In Figure~\ref{fig:SECHD} we see a familiar increase in opacity and change in morphology for the cross-section method. The pressure-broadening is also slightly more significant compared to primary transit, as we are probing lower in the atmosphere. The effects here are smaller than for the brown dwarfs, where a change in the method can increase the flux by half an order of magnitude in the mid- and far- infrared, or change it by an order of magnitude in the near-IR. The effects of pressure-broadening in secondary eclipse is smaller than on brown dwarfs (because the contribution functions peak at a lower pressure in the atmosphere) but still appreciable enough to become apparent in the spectra. 

\section{Conclusions}

A major discrepancy between different radiative transfer models is how gaseous absorption is calculated. We investigated the effects of using two different gaseous absorption methods (correlated-$k$ and cross-sections) to calculate spectra in a variety of atmospheres of varying complexities. We also investigated the effects of including \ce{H2}-He pressure broadening in the more complex atmospheres. These investigations are important because radiative transfer models are often coupled to inverse methods, which can iterate over millions of forward models in a large parameter space; a single forward model error, such as a `ghost feature', would propagate through to all retrieved PDFs influencing them in a non-trivial way, i.e. an error in the absorption spectrum of \ce{NH3} could influence the retrieved temperature profile.

We first showed that for test cases with resolutions of $\Delta \nu = 1$cm$^{-1}$ the cross-section method overestimates the amount of absorption present in the atmosphere and should therefore be used with caution. The morphology of the spectra also changes and produces `ghost' features for mixed-gas atmospheres. When considering our flattened $k$-table cross-sections, the effect can produce multiple orders of magnitude change in the flux received from brown dwarfs in certain wavelength regions. The effect is similar but smaller for primary transit, and is closer in order to a $\sim$ 1\% change in transit depth. The flux ratio of secondary eclipse exoplanets can find an order of magnitude change in the near-IR, with the effect becoming lesser for longer wavelengths. Correlated-$k$ can produce similar results to line-by-line and very high-resolution cross-sections, but is much less computationally expensive.
The inclusion of \ce{H2}-He pressure broadening similarly changes the total flux found in the spectra of brown dwarfs and secondary transit exoplanets by up to an order of magnitude, while only making slight changes to the spectra of transiting exoplanets. 

If we take into account the sizable discrepancies found by \citet{hedges2016} between the different aspects of pressure broadening from medium- and higher resolutions, the issues discussed in this paper might be even more serious than suggested. However, at higher resolutions the differences we present here will become less significant, and so the main error source will switch from those discussed here to those discussed by \citet{hedges2016}.

We conclude that inaccurate use of cross-sections and omission of pressure broadening can be key sources of error in the modelling of brown dwarf and exoplanet atmospheres. These sources of error in the forward model may produce strong biases in the probability distribution functions of retrieved parameters. 

\section*{Acknowledgements}

R.G. thanks and acknowledges the support of the Science and Technology Facilities Council.  P.G.J.I. also receives funding from the Science and Technology Facilities Council (ST/K00106X/1).




\bibliographystyle{mnras}
\bibliography{master} 


\newpage
\appendix
\newpage

\begin{landscape}
\begin{table}
\caption{Summary of Line Lists Selected. Data sources: HITEMP: \url{http://www.cfa.harvard.edu/hitran/HITEMP.html}, CDSD:  \url{ftp://ftp.iao.ru/pub/CDSD-4000}, ExoMol: \url{www.exomol.com}, VALD: \url{http://vald.astro.uu.se}.} 
\label{tbl:linelists}
\begin{tabular}{cccc}
Molecule & Reference & Q(T)& Available \\
\hline

CO & \citet{rothman10} &  \citet{rothman10} & HITEMP \\
\ce{CO2} & \citet{tashkun11} & \citet{rothman13} & CDSD  \\ 
\ce{H2O} & \citet{barber06} & \citet{barber06} & ExoMol \\
\ce{NH3} & \citet{yurchenko11} & \citet{yurchenko11} & ExoMol \\
\ce{CH4} & \citet{yurchenko14} & \citet{yurchenko14} & ExoMol \\
TiO & R.S. Freedman (priv. com.) & \citet{sauval84} & N/A \\
VO & R.S. Freedman (priv. com.) & \citet{sauval84} &N/A \\
Na & \citet{kupka00} & \citet{sauval84} & VALD \\
K & \citet{kupka00} & \citet{sauval84} & VALD \\ 

\end{tabular}
\end{table}
\end{landscape}

\begin{landscape}
\begin{table}
\caption{Summary of Line Broadening Data}
\begin{tabular}{ccc|c|c}
Molecule & Broadener & Line width reference & $n$ (Average value) & Temp. exponent reference \\
\hline
\multirow{2}{*}{\ce{H2O}} & \ce{H2} & \citet{gamache96} & 0.44 & \citet{gamache96} \\ & He & \citet{solodov09,steyert04} & 0.44 & \citet{gamache96} \\
\hline
\multirow{2}{*}{\ce{CH4}} & \ce{H2} & \citet{pine92,margolis93} & 0.44 & \citet{margolis93}\\ & He & \citet{pine92} & 0.28 & \citet{varanasi90} \\
\hline
\multirow{2}{*}{\ce{CO2}} & \ce{H2} & \citet{padmanabhan14} & 0.60 & \citet{sharp07} \\ & He &  \citet{thibault92} & 0.60 & \citet{thibault00} \\
\hline
\multirow{2}{*}{\ce{CO}} & \ce{H2} & \citet{regaliajarlot05} & 0.60 & \citet{moal86} \\ & He & \citet{belbruno82, mantz05} & 0.55 & \citet{mantz05} \\
\hline
\multirow{2}{*}{\ce{NH3}} & \ce{H2} & \citet{hadded01,pine93} & 0.64 & \citet{nouri04} \\ & He & \citet{hadded01,pine93} & 0.40 & \citet{sharp07} \\
\hline
\multirow{2}{*}{\ce{TiO}} & \ce{H2} & \citet{sharp07} & 0.60 & \citet{sharp07} \\ & He & \citet{sharp07} & 0.40 & \citet{sharp07} \\
\hline
\multirow{2}{*}{\ce{VO}} & \ce{H2} & \citet{sharp07} & 0.60 & \citet{sharp07} \\ & He & \citet{sharp07} & 0.40 & \citet{sharp07} \\
\end{tabular}
\label{tbl:presbroad}
\end{table}
\end{landscape}


\label{lastpage}
\end{document}